\documentclass[a4paper,11pt]{article}
\usepackage{jheppub} 
\usepackage{lineno}

\arxivnumber{2504.20646} 

\title{\boldmath From the Fokker-Planck equation to perturbative QFT's results in de Sitter space}

\author[]{Alexander Kamenshchik} \author[]{and Polina Petriakova} 
\affiliation[]{Dipartimento di Fisica e Astronomia, Universit\`{a} di Bologna, \\
via Irnerio 46, 40126 Bologna, Italy}
\affiliation[]{I.N.F.N., Sezione di Bologna, I.S. FLAG, \\
viale B. Pichat 6/2, 40127 Bologna, Italy}

\emailAdd{kamenshchik@bo.infn.it} 
\emailAdd{polina.petriakova@bo.infn.it}

\abstract{A technique to build perturbative series for the spectator field's correlation functions in de Sitter space through the Fokker-Planck equation is proposed. We derive from the first-order differential equation the iterative integral relation for equal- and multi-time correlation functions. With an appropriate integration, the obtained two-point and four-point correlation functions are in agreement with quantum field theory's ones.}

\begin{document}
\maketitle
\flushbottom
\section{{Introduction}}\label{Introduction}
The description of the quantum scalar field's infrared part in de Sitter space through the classical stochastic field was invented by Starobinsky~\cite{1986LNP...246..107S}. One decomposes the operator of the quantum field $\phi\,(t,\x)$ in the Heisenberg representation into the super-Hubble or long-wavelength, $H<k<H\e^{Ht}$, and the sub-Hubble or short-wavelength, $k>H\e^{Ht}$, parts by the window function, to be chosen by the dynamical Heaviside step function. The long-wavelength part $\phi^{\text{l-w}}(t,\x )$ satisfies the equation of motion $ \Box \, \phi=-V^{\,\prime}_{\phi} \bigl( \phi \bigr)$, and for the slowly varying $\phi^{\text{l-w}}(t,\x )$, it takes the local Langevin-like form~\cite{1986LNP...246..107S}, where the noise term represents short-wavelength modes that continually shift into the long-wavelength ones. Therefore, the long-wavelength part of the quantum scalar field $\phi^{\text{l-w}}(t,\x )$ can be treated as the classical stochastic field~$\varphi(t,\x)$ 
with a probability distribution function $\mathcal{P} \bigl[\varphi(t,\x)\bigr]$ that satisfies the Fokker-Planck or Einstein-Smoluchowski equation, associated to the corresponding local Langevin-like equation. Later on, the stationary solution to the Fokker-Planck equation and the related method to calculate equal-time correlation functions at the equilibrium were presented by Starobinsky and Yokoyama~\cite{1994PhRvD..50.6357S}. The so-called stochastic treatment efficiently catches the leading infrared logarithms in scalar field models. Moreover, the resummed non-perturbative result is free of secular growth, which is present for the massless scalar field’s two-point correlation function already at the tree level~\cite{Vilenkin:1982wt,Linde:1982uu}.

The program to establish the correspondence between the stochastic approach and the results from perturbative quantum field theory in the leading logarithm approximation was launched by Woodard and collaborators~\cite{2005NuPhS.148..108W,2005NuPhB.724..295T}; see the recent one~\cite{2025arXiv250115843W} for the full reference~list of further developments by that group and by many others~\cite{2007PhRvD..76d3512K,2013arXiv1305.0229R,2014PhRvD..89f3506G,2015PhRvD..91f3520G,2015PhRvD..91j3537O,2017JHEP...11..065C,2018PhRvD..97l3531K,2020JHEP...07..119B,2020PhRvD.102f5010K,2022EPJC...82..345K,2022JHEP...02..121A,2023JCAP...03..058B,2023JHEP...08..060H,2024JHEP...04..004C}. To trace it out, authors of~\cite{2005NuPhS.148..108W,2005NuPhB.724..295T} considered a massless scalar field with the quartic self-interaction and retrieved the stochastic expectation value's series through the Fokker-Planck equation with the help of some recurrent relations. The extracted series turned out to be in agreement with ones from the quantum field theory~\cite{1998PhLB..426...21T,2007PhRvD..76d3512K,2015PhRvD..91j3537O,2020PhRvD.102f5010K}. In addition, Tsamis and Woodard derived the truncated Yang-Feldman equation~\cite{2005NuPhS.148..108W,2005NuPhB.724..295T,Yang:1950vi}:
\begin{equation}\label{Y-F_k0}
\phi\left(t, \x \right) = \phi_0\left(t, \x \right) -\frac{\lambda}{3H} \int\limits_{0}^{t} dt^\prime \, 
 \phi^3\left(t^\prime, \x \right).
\end{equation}
That equation recursively defines the interacting scalar field $\phi\left(t, \x \right)$ as a self-interaction coupling constant's formal power series through the free $\phi_0\left(t, \x \right)$ one. 
By iterating~\eqref{Y-F_k0} up to the necessary $\lambda$ order, one builds up the formal perturbative series for the Heisenberg operator and computes the vacuum expectation value for the operator's product, which is in agreement with the diagrammatic QFT results~\cite{1998PhLB..426...21T,2005NuPhB.724..295T,2007PhRvD..76d3512K,2015PhRvD..91j3537O,2020PhRvD.102f5010K} in the leading logarithm approximation. Remarkably, by taking the time derivative of~\eqref{Y-F_k0}, it precisely coincides~\cite{2005NuPhS.148..108W,2005NuPhB.724..295T} with the local Langevin-like equation for the stochastic field in the original work~\cite{1986LNP...246..107S}.

In our previous article~\cite{2025JHEP...04..127K}, we investigated the two-point and four-point correlation functions in the long-wavelength approximation within a particular theory of a minimally coupled massive scalar field living on the flat de Sitter background
\begin{equation}\label{S0}
\mathcal{L}_m = \sqrt{-g} \, \left( \frac12 \, \phi^{\, , \mu}\phi_{, \mu} - \frac12 m^2 \phi^2 - \frac{\lambda}{4} \, \phi^4 \right), \qquad ds^2=dt^2- \e^{2Ht}d \x^2.
\end{equation}
We proposed a method to include the mass in the Yang-Feldman equation~\eqref{Y-F_k0} as follows: 
\begin{equation}\label{Y-F-massive}
 \phi\left(t, \x \right) = \tilde{\phi}\left(t, \x \right) - \frac{\lambda}{3H} \, \e^{-\tfrac{m^2t}{3H}} \int\limits_{0}^{t} dt^\prime \, \e^{\tfrac{m^2t^\prime}{3H}} \, {\phi}^{\,3}\left(t^\prime, \x \right),
\end{equation} where the free massless scalar field $\phi_0\left(t, \x \right)$ is replaced by the free massive one $\tilde{\phi}\left(t, \x \right)$, and they are related through the corresponding integral equation:
\begin{equation}\label{FreeMassiveField}
\tilde{\phi}\left(t, \x \right)  = \phi_0 \left(t, \x \right) - \frac{m^2}{3H} \, \e^{-\tfrac{m^2t}{3H}} \int\limits_{0}^{t} dt^\prime \, \e^{\tfrac{m^2t^\prime}{3H}} \, \phi_0 \left(t^\prime, \x \right). \end{equation}
Those equations permit the vacuum expectation value's computation of the free massive scalar field and the quantum corrections relying only on the vacuum expectation value of the free massless one. The retarded Green's function under the integral acquires the additional exponent factor in~\eqref{Y-F-massive} in the massive case; the same retarded Green's function form is extracted in~\cite{2013arXiv1305.0229R,2014PhRvD..89f3506G,2015PhRvD..91f3520G}. The proposed treatment in~\cite{2025JHEP...04..127K} is equivalent to the following: one builds up the perturbative series on $m^2$ for the massless scalar field, sums up the leading logarithms (thanks to the massive theory to be the Gaussian), and then organizes the perturbative massive series on~$\lambda$. Within the truncated Yang-Feldman-type equation~\eqref{Y-F-massive}, our "building block" is the free massive field's 
vacuum expectation value, derived via~\eqref{FreeMassiveField}:
\begin{equation}\label{O-U_correlator_QFT}
\Bigl\langle \tilde{\phi} (t_1) \, \tilde{\phi} (t_2) 
\Bigr \rangle = \frac{3H^4}{8 \, \pi^2m^2} \, \biggl( \e^{-\tfrac{m^2}{3H}|t_1-t_2|} - \e^{-\tfrac{m^2}{3H}(t_1+t_2)} \biggr).  
\end{equation}
It coincides with the well-known Ornstein-Uhlenbeck correlation function~\cite{Uhlenbeck:1930zz,1989fpem.book.....R} and lacks de Sitter invariance. By its virtue, the obtained correlation function both has the smooth massless limit and tends to the equilibrium, de Sitter-invariant state over time, as well as the perturbative series, organized through~\eqref{Y-F-massive}. That expression~\eqref{O-U_correlator_QFT} was also obtained in~\cite{2014PhRvD..89f3506G,2015PhRvD..91f3520G}, while the attractor feature to the de Sitter-invariant state was discovered in the preceding works~\cite{1994PhRvD..50.6357S,2000PhRvD..62l4019A} from different perspectives. We revealed in~\cite{2025JHEP...04..127K} that the two-point correlation function~\eqref{O-U_correlator_QFT} is the resummed perturbative series on $m^2$ within the leading infrared logarithm approximation. Furthermore, we have established the correspondence of the integral structures arising in the iterative Yang-Feldman-type equation~\eqref{Y-F-massive} and diagrams within the Schwinger-Keldysh (or "in{--}in") formalism. Our series for the two-point correlation function at late times is in agreement with the direct "in{--}in" ones~up to the two-loop order~\cite{2013PhLB..727..541G,2022EPJC...82..345K}. In the case of equal-time correlation functions at the equilibrium, our outcomes match those of the Starobinsky stochastic approach~\cite{1986LNP...246..107S,1994PhRvD..50.6357S} in the perturbative regime and those obtained in~\cite{2014PhRvD..89f3506G,2015PhRvD..91f3520G} at the late-time limit. Indeed,~the truncated Yang-Feldman-type equation in the massive case~\eqref{Y-F-massive} coincides with Starobinsky's Langevin equation too. By~taking the time derivative of~\eqref{Y-F-massive} with the use of auxiliary equation~\eqref{FreeMassiveField}, one gets the same equation form: the local Langevin-like one from~\cite{1986LNP...246..107S}.

The present work provides an alternative way to compute the perturbative series for equal-time and multi-time correlation functions. 
Instead of processing the Yang-Feldman-type equation or dealing with the diagrammatic techniques, one derives a simple first-order differential equation via the Fokker-Planck or forward Kolmogorov equation. Its~formal solution relates various correlation functions at different self-interaction coupling constant~$\lambda$ orders. Two-point and four-point correlation functions, obtained as a result of appropriate integration, are in agreement with those by quantum field theory's methods.

This paper is organized as follows: in section~\ref{Section_equal} we show how to deal with the equal-time correlation functions. In the equal-time case, one relies on the straightforward integration of the corresponding integral relation between various correlation functions at the different~$\lambda$ orders. In section~\ref{Section_multi} we extend our approach to the multi-time correlation functions and reveal the appropriate treatment there. Section~\ref{Conclusion} contains our summary.

\section{{Equal-time correlation functions}}\label{Section_equal}
In this section, we present an approach to the computation of the equal-time correlation functions directly through the Fokker-Planck equation.

The corresponding Fokker-Planck or forward Kolmogorov equation for our model under consideration reads~\cite{1986LNP...246..107S,1994PhRvD..50.6357S} 
\begin{equation}\label{Fokker-Planck}
    \frac{\partial}{\partial t} \, \mathcal{P} \bigl[\varphi \, ( t \, , \x )\bigr] = \frac{1}{3H} \, \frac{\partial}{\partial \varphi} \, \Bigl( V^\prime_\varphi \bigl(\varphi\, ( t \, , \x )\bigr) \, \mathcal{P} \bigl[\varphi \, ( t \, , \x )\bigr] \Bigr) + \frac{H^3}{8\pi^2} \frac{\partial^2}{\partial \varphi^2} \, \mathcal{P} \bigl[\varphi \, ( t \, , \x )\bigr]. 
\end{equation}
Here, $V^\prime_\varphi \bigl(\varphi\, ( t \, , \x )\bigr) $ is the same potential as one studies within the quantum field theory; in our setup that is~\eqref{S0}. One can multiply both sides by ${\varphi}^{2\n}$ and integrate over $\int\limits_{-\infty}^{+\infty} d {\varphi}$ with the probability density $\mathcal{P} \, \bigl[{\varphi}(t,\x)\bigr]$, 
and after integration by parts, it results in
\begin{equation}\label{diff_first-order_FPK}
\frac{\partial}{\partial t} \Bigl\langle \varphi^{2\n} (t,\x) \Bigr \rangle  = - \frac{2 \n m^2}{3H} \, \Bigl\langle \varphi^{2\n} (t,\x) \Bigr \rangle - \frac{2 \n\lambda}{3H} \, \Bigl\langle \varphi^{2\n+2} (t,\x) \Bigr \rangle + \n \, \bigl( 2\n - 1 \bigr) \, \frac{H^3}{4\pi^2} \, \Bigl\langle \varphi^{2\n-2} (t,\x) \Bigr \rangle.
\end{equation} 
Let us point out that this system of equations must be valid for any value of $\n$, while in the literature one can find the explored equation after the multiplication on $\varphi^2$ only~\cite{2017JCAP...10..018H}. Hereafter, we omit the argument $\x$ since it is the same in all our expressions. The differential equation above can be easily rewritten in the following form:
\begin{equation}
\e^{- \tfrac{2\n m^2 t}{3H}}\frac{\partial}{\partial t}\biggl( 
\e^{\tfrac{2\n m^2 t}{3H}} \Bigl\langle \varphi^{2\n} (t)
\Bigr \rangle 
\biggr) = - \frac{2 \n\lambda}{3H} \, \Bigl\langle \varphi^{2\n+2} (t)
\Bigr \rangle + \n \, \bigl( 2\n - 1 \bigr) \, \frac{H^3}{4\pi^2} \, \Bigl\langle \varphi^{2\n-2} (t)
\Bigr \rangle, 
\end{equation}
and, therefore, we have the integral relation for equal-time correlation functions
\begin{equation}\label{Fokker-Planck_gen}
\Bigl\langle \varphi^{2\n} (t) \Bigr \rangle 
= \e^{-\tfrac{2\n m^2 t}{3H}} \int\limits_0^t dt^\prime \, \e^{\tfrac{2\n m^2 t^\prime}{3H}} \biggl(- \frac{2 \n\lambda}{3H} \, \Bigl\langle \varphi^{2\n+2} (t^\prime) \Bigr \rangle + \, \n \, \bigl( 2\n - 1 \bigr) \frac{H^3}{4\pi^2} \, \Bigl\langle \varphi^{2\n-2} (t^\prime) \Bigr \rangle  \biggr).    
\end{equation}
The obtained relation connects various correlation functions at the different self-interaction coupling constant~$\lambda$ orders. 

The starting point is to define equal-time correlation functions in the free massive case without self-interaction, $\lambda=0$. In that case, one finds an exact solution to the Fokker-Planck equation~\eqref{Fokker-Planck} for the probability distribution function $\mathcal{P}_m \bigl[\varphi(t,\x)\bigr]$. By taking the following ansatz
\begin{equation}\label{ansatz_massive}
\mathcal{P}_m \bigl[\varphi(t,\x)\bigr] = \frac{\e^{-\varphi^2/ \, \zeta(t)}}{\sqrt{\pi \zeta (t) \,}} 
\end{equation} and substituting it into~\eqref{Fokker-Planck}, one gets a simple equation for the unknown function and the corresponding solution
\begin{equation}\label{eq_for_ansatz}
\dot{\zeta} = -\frac{2m^2}{3H} \, \zeta(t) + \frac{H^3}{2\pi^2} \qquad \Rightarrow \qquad \zeta(t) = \frac{3H^4}{4\, \pi^2m^2} \biggl( 1 - \e^{-\tfrac{2m^2t}{3H}} \biggr).
\end{equation}
With the help of the obtained solution, we can calculate any $2\n$’th correlation functions as
\begin{equation}\label{2pt_massive_free}
\Bigl \langle \varphi^{2\n}(t) \Bigr \rangle = \bigl(2\n-1\bigr)!! \,\, \frac{\zeta^\n(t) }{2^{\, \n}} \equiv \Bigl \langle \varphi^{2\n}(t) \Bigr \rangle_0.
\end{equation} 

One organizes the iterative series $\bigl \langle \varphi^{2\n}(t) \bigr \rangle = \sum\limits_{\kk=0}^\infty \,\, \lambda^\kk \, \bigl \langle \varphi^{2\n}(t) \bigr \rangle_\kk$ 
and order by order gains the terms through~\eqref{Fokker-Planck_gen}:
\begin{align}\label{reccurent_equation_massive_and_lambda}
\Bigl\langle \varphi^{2\n} (t) \Bigr \rangle_{\kk+1} 
= \e^{-\tfrac{2\n m^2 t}{3H}} \int\limits_0^t \, dt^\prime \, \e^{\tfrac{2\n m^2 t^\prime}{3H}} \, \biggl( & - \, \frac{2 \n\lambda}{3H} \, \Bigl\langle \varphi^{2\n+2} (t^\prime) \Bigr \rangle_{\kk} \\ \nonumber 
& \, + \, \n \, \bigl( 2\n - 1 \bigr) \, \frac{H^3}{4\pi^2} \, \Bigl\langle \varphi^{2\n-2} (t^\prime) \Bigr \rangle_{\kk+1} \, \biggr).    
\end{align}
We rely on this interconnection to derive various correlation functions and to compare them with ones found in our previous work~\cite{2025JHEP...04..127K} and other equilibrium results in the literature. 

Let us start from the two-point correlation function. In that case, according to~\eqref{reccurent_equation_massive_and_lambda}, on the $\lambda${--}linear order, we have to integrate the "zero"th order four-point correlation function. 
Hereafter, all correlation functions indexed by zero we relate to the free massive case and present them in~\eqref{2pt_massive_free}. Besides, $\bigl\langle \varphi^{0} (t) \bigr \rangle= 1$; it does not contribute to the $\lambda${--}linear order~in that case. With the integration of $\bigl\langle \varphi^{0} (t) \bigr \rangle= 1$, $\bigl\langle \varphi^{2} (t) \bigr \rangle$, etc. in~\eqref{Fokker-Planck_gen} in the free massive case, one obtains exactly~\eqref{2pt_massive_free}. Within the same equation~\eqref{reccurent_equation_massive_and_lambda}, we are able to compute the four-point correlation function at the linear $\lambda$ order: one should, obviously, integrate out the six-point correlation function at the zero order, $\bigl\langle \varphi^{6} (t) \bigr \rangle_{0}$, from~\eqref{2pt_massive_free} and the two-point one at the $\lambda${--}linear order, which one knows from the previous step. Next, we find the two-point correlation function at the $\lambda^2$ order, again, through~\eqref{reccurent_equation_massive_and_lambda}. In that fashion, for the $\lambda^2$ order of the four-point correlation function, we compute the $\lambda${--}linear order for $\bigl\langle \varphi^{6} (t) \bigr \rangle_\lambda$, and, finally, with the appropriate integration, one obtains $\bigl\langle \varphi^{2} (t) \bigr \rangle_{\lambda^3}$. Below we are listing the final results of that trivial integration performance:
\begin{align}\label{2pt_massive_lambda_series}
& \Bigl\langle \varphi^{2} (t) \Bigr \rangle = \frac{3H^4}{8\, \pi^2m^2} \biggl( 1 - \e^{-\tfrac{2m^2t}{3H}} \biggr)  - \frac{27 \, \lambda H^8}{64 \, \pi^4 m^6}  \biggl( 1 - \frac{4m^2 t}{3H}\, \e^{-\tfrac{2 m^2 t}{3H}} - \e^{-\tfrac{4 m^2 t}{3H}} \biggr) \\ \nonumber 
& 
\,\,\,\, + \, \frac{81 \lambda^2 H^{12}}{64 \, \pi^6 m^{10}} \Biggl( 1 + \biggl( \frac{21}{8} - \frac{3m^2t}{2H} - \frac{m^4t^2}{3H^2} \biggr) \, \e^{-\tfrac{2 m^2 t}{3H}} 
- \biggl( 3 + \frac{2m^2t}{H} \biggr) \, \e^{-\tfrac{4 m^2 t}{3H}} - \frac{5}{8} \, \e^{-\tfrac{6 m^2 t}{3H}}\Biggr) \\ \nonumber 
& 
\,\,\,\, - \, \frac{2187 \lambda^3 H^{16}}{4096 \, \pi^8 m^{14}} \Biggl( 11 + \biggl( \frac{872}{9} - \frac{172\,m^2t}{9H} - \frac{16m^4t^2}{3H^2} - \frac{32m^6t^3}{81H^3} \biggr) \, \e^{-\tfrac{2 m^2 t}{3H}} \\ \nonumber 
& 
\,\,\,\, - \biggl( 64 + \frac{72 \, m^2t}{H} + \frac{128 \, m^4t^2}{9H^2}\biggr) \, \e^{-\tfrac{4 m^2 t}{3H}}  - \biggl( 40 + \frac{20\,m^2t}{H} \biggr) \, \e^{-\tfrac{6m^2 t}{3H}} - \frac{35}{9} \, \e^{-\tfrac{8 m^2 t}{3H}}\Biggr) + O\left(\lambda^4\right);  \\ 
\label{4pt_massive_lambda_series}
& \Bigl\langle \varphi^{4} (t) \Bigr \rangle = \frac{27H^8}{64 \, \pi^4m^4} \, \biggl( 1 - \e^{-\tfrac{2m^2 t}{3H} } \biggr)^2 \, - \, \frac{81 \lambda H^{12}}{64 \, \pi^6 m^8} \Biggl(1- \left(\frac{9}{4} + \frac{m^2 t}{H} \right) \e^{-\tfrac{2m^2 t}{3H}} \\ \nonumber 
& \, + \frac{2m^2t}{H} \e^{-\tfrac{4m^2 t}{3H}} + \frac{5}{4} \e^{-\tfrac{2m^2 t}{H}}\Biggr) + \frac{729 \lambda^2 H^{16}}{4096 \, \pi^8 m^{12}} \Biggl( \, 33 \, - \biggl( 86 + \frac{48m^2 t}{H} + \frac{16 m^4 t^2}{3H^2}\biggr) \e^{-\tfrac{2m^2t}{3H}} \\ \nonumber 
& \, - \biggl( 132 - \frac{88m^2 t}{H}  - \frac{128m^4 t^2}{3H^2}\biggr) \e^{-\tfrac{4m^2t}{3H}} + \biggl( 150 + \frac{120m^2 t}{H} \biggr) \e^{-\tfrac{2m^2t}{H}} + 35 \, \e^{-\tfrac{8m^2t}{3H}}\Biggr)  + O\left(\lambda^3\right); \\ \label{6pt_massive_lambda_series}
&  \Bigl\langle \varphi^{6} (t) \Bigr \rangle = \frac{405H^{12}}{512 \, \pi^6 m^6} \, \biggl( 1 - \e^{-\tfrac{2m^2 t}{3H} } \biggr)^3 \, - \frac{18225 \, \lambda H^{16}}{4096 \, \pi^8 m^{10}} \Biggl( 1 \, - \biggl( 4 + \frac{4m^2t}{5H} \biggr) \, \e^{-\tfrac{2 m^2 t}{3H}} \\ \nonumber 
& \qquad \qquad \qquad \,\,\,\, + \biggl( \frac{18}{5} + \frac{16m^2t}{5H} \biggr) \, \e^{-\tfrac{4 m^2 t}{3H}} + \, \biggl( \frac{4}{5} - \frac{12m^2t}{5H} \biggr) \, \e^{-\tfrac{6 m^2 t}{3H}} -\frac{7}{5} \, \e^{-\tfrac{8 m^2 t}{3H}}\Biggr) + O\left(\lambda^2\right).
\end{align}
All of them, namely, $\lambda$, $\lambda^2$, and $\lambda^3$ orders for the two-point and $\lambda$ and $\lambda^2$ for the four-point correlation functions, coincide with our outcomes from the previous work~\cite{2025JHEP...04..127K}, which were obtained by processing the Yang-Feldman-type equation~\eqref{Y-F-massive}. Our massive perturbative on $\lambda$ series with $m^2 \ll H^2$ and $\lambda \ll m^4/H^4$ converge and, at the late-time limit, reach
\begin{align}\label{two-point_equilibrium}
\Bigl\langle \varphi^{2}(t) \Bigr \rangle
& \xrightarrow[\,\, \text{times}\,\,]{\text{late}} \, \frac{3H^4}{8\, \pi^2m^2} - \, \frac{27 \, \lambda H^8}{64 \, \pi^4 m^6} + \, \frac{81 \lambda^2 H^{12}}{64 \, \pi^6 m^{10}} - \, \frac{24057 \lambda^3 H^{16}}{4096 \, \pi^8 m^{14}} + \, O\left(\lambda^4\right); \\ \label{four-point_equilibrium}
\Bigl\langle \varphi^{4} (t) \Bigr \rangle
& \xrightarrow[\,\, \text{times}\,\,]{\text{late}} \, \frac{27H^8}{64 \, \pi^4m^4} - \, \frac{81 \lambda H^{12}}{64 \, \pi^6 m^8} + \, \frac{24057 \lambda^2 H^{16}}{4096 \, \pi^8 m^{12}} + \, O\left(\lambda^3\right);
\\ \label{six-point_equilibrium}
\Bigl\langle \varphi^{6} (t) \Bigr \rangle
& \xrightarrow[\,\, \text{times} \,\, ]{\text{late}} \, \frac{405H^{12}}{512 \, \pi^6 m^6} - \, \frac{18225 \, \lambda H^{16}}{4096 \, \pi^8 m^{10}} + \, O\left(\lambda^2\right).
\end{align} On the expansion scheme, we provide some details in Appendix~\ref{Leading_log}. Our results~\eqref{2pt_massive_lambda_series}{--}\eqref{6pt_massive_lambda_series} are equivalent to the resummation of leading infrared logarithms in the perturbative on $m^2$ series at each $\lambda$ order. That is precisely what we revealed in the previous work~\cite{2025JHEP...04..127K} and discussed in the introduction; see~\eqref{Y-F-massive}{--}\eqref{O-U_correlator_QFT} and nearby reasoning. Within this scheme the massless time divergences of the self-interacting scalar field are fully resummed. The obtained two-point correlation function's perturbative series at the equilibrium~\eqref{two-point_equilibrium} is in agreement with one from~\cite{2014PhRvD..89f3506G,2015PhRvD..91f3520G,2022EPJC...82..345K} up to $O\left(\lambda^3\right)$. In addition, outcomes~\eqref{two-point_equilibrium}{--}\eqref{six-point_equilibrium} match the non-perturbative $\bigl\langle \varphi^{2\n} \bigr \rangle^{\text{late-time}}$ from~\cite{2025JHEP...04..127K} in the perturbative on $\lambda$ regime; we derived it through some recurrence relations within Starobinsky's stochastic approach~\cite{1986LNP...246..107S,1994PhRvD..50.6357S}.

Regarding the massless case, our approach is also clearly efficient. In this case, the ansatz has the same form~\eqref{ansatz_massive}, and the equation for the unknown function is~\eqref{eq_for_ansatz} without the massive term. The solution to the corresponding equation is $\zeta_{m=0}(t)=H^3t/2\pi^2$ and coincides with the smooth massless limit of the solution in~\eqref{eq_for_ansatz}, as one expects~\cite{Linde:1982uu,Vilenkin:1982wt}. The $2\n$'th correlation function in the case without self-interaction is defined by the identical-to~\eqref{2pt_massive_free} relation
\begin{equation}\label{zero_massless}
\Bigl \langle \varphi^{2\n}_{m=0}(t) \Bigr \rangle_0 = \bigl(2\n-1\bigr)!! \,\, \left(\frac{H^3t}{4\pi^2}\right)^\n ,  
\end{equation} and the resulting formal solution to the corresponding first-order differential equation~\eqref{diff_first-order_FPK} without the massive term takes form~\eqref{reccurent_equation_massive_and_lambda} without exponents:
\begin{equation}\label{reccurent_equation_massless_and_lambda}
\Bigl\langle \varphi^{2\n}_{m=0} (t) \Bigr \rangle_{\kk+1} 
= \int\limits_0^t \, dt^\prime \,  \biggl(- \, \frac{2 \n\lambda}{3H} \, \Bigl\langle \varphi^{2\n+2}_{m=0} (t^\prime) \Bigr \rangle_{\kk} + \, \n \, \bigl( 2\n - 1 \bigr) \, \frac{H^3}{4\pi^2} \, \Bigl\langle \varphi^{2\n-2}_{m=0} (t^\prime) \Bigr \rangle_{\kk+1} \, \biggr).
\end{equation} Let us also note that in our previous article~\cite{2025JHEP...04..127K} in the massive case, the Yang-Feldman-type equation (or the retarded Green's function; see also~\cite{2013arXiv1305.0229R,2014PhRvD..89f3506G,2015PhRvD..91f3520G}) acquires the additional exponents in the same spirit in comparison to the massless one. With the use of~\eqref{zero_massless} and~\eqref{reccurent_equation_massless_and_lambda} and following the proposed routine, one calculates the series
\begin{align}\label{2pt_massless_series}
\Bigl\langle \varphi^2_{m=0} \left(t \right) \Bigr\rangle & = 
\, \frac{H^3t}{4\pi^2} - \frac{\lambda H^5 t^3}{24 \, \pi^4} + \frac{\lambda^2 H^7 t^5}{80 \, \pi^6} - \frac{53 \lambda^3 H^9 t^7}{10080 \, \pi^8} + O\left(\lambda^4\right); \\ \label{4pt_massless_series}
\Bigl\langle \varphi^4_{m=0} \left(t \right) \Bigr\rangle & =
\frac{3 H^6 t^2}{16 \, \pi^4} -   \frac{3 \lambda H^8 t^4}{32 \, \pi^6} +  \frac{53 \, \lambda^2 H^{10} \, t^6}{960 \, \pi^8} + O\left(\lambda^3\right); \\ \label{6pt_massless_series}
\Bigl\langle \varphi^6_{m=0} \left(t \right) \Bigr\rangle & =
\frac{15 H^9 t^3}{64 \, \pi^6} - \frac{15 \lambda H^{11} t^5}{64 \, \pi^8} + O\left(\lambda^2\right),
\end{align} which are precisely the massless limits of~\eqref{2pt_massive_lambda_series},~\eqref{4pt_massive_lambda_series}, and~\eqref{6pt_massive_lambda_series}. Outcomes also coincide with the corresponding results up to some (which are present in the references onward) orders within the quantum field theory tools~\cite{1998PhLB..426...21T,2007PhRvD..76d3512K,2015PhRvD..91j3537O,2020PhRvD.102f5010K,2025JHEP...04..127K} and are in agreement with the Starobinsky stochastic approach~\cite{2005NuPhB.724..295T,2025JHEP...04..127K} in the perturbative regime. One notices the recovering of the leading infrared logarithm structure there in the massless limit.

\section{{Multi-time correlation functions}}\label{Section_multi} 

Now we are in a position to generalize our approach to the multi-time correlation functions. The idea is the same: to derive the first-order differential equation from the Fokker-Planck equation and the integral relation for the multi-time correlation functions through it.

In that case, the conditional (or transition) probability $\mathcal{P} \, \bigl(\, \varphi, t \, \bigl| \, \varphi_0, t_0 \bigr)$ must also obey the Fokker-Planck equation or the forward Kolmogorov equation for $t \geq t_0$:
\begin{equation}\label{Fokker-Planck_General_t1t2}
    \frac{\partial}{\partial t} \, \mathcal{P} \bigl(\, \varphi, t \, \bigl| \, \varphi_0, t_0 \bigr) = \frac{1}{3H} \, \frac{\partial}{\partial \varphi} \, 
    \biggl( \Bigl(m^2 \varphi + \lambda \varphi^3 \Bigr) \, \mathcal{P} \bigl(\, \varphi, t \, \bigl| \, \varphi_0, t_0 \bigr) \biggr)  
    + \frac{H^3}{8\pi^2} \frac{\partial^2}{\partial \varphi^2} \, \mathcal{P} \, \bigl(\, \varphi, t \, \bigl| \, \varphi_0, t_0 \bigr) 
\end{equation}
and have the initial value $ \mathcal{P} \bigl(\varphi, t_0 \, \bigl| \, \varphi_0, t_0 \bigr)=\delta\bigl(\varphi-\varphi_0\bigr)$.

For the massive case without self-interaction, namely, $\lambda=0$, the exact solution to~\eqref{Fokker-Planck_General_t1t2} for the conditional (or transition) probability is known:
\begin{equation} 
\label{sol_prob_exact_massive}
\mathcal{P}_m \bigl( \varphi, t \bigl| \, \varphi_0, t_0 \bigr) = \frac{2\sqrt{\pi} m}{\sqrt{3} \, H^2} \biggl(1 - \,\e^{-\tfrac{2m^2}{3H}(t-t_0)}\biggr)^{-1/2}
\text{exp}\left(-\frac{4\pi^2m^2}{3H^4} \frac{\Bigl( \varphi - \varphi_0 \, \e^{-\tfrac{m^2}{3H}(t-t_0)}\Bigr)^2}{\Bigl(1-\e^{-\tfrac{2m^2}{3H}(t-t_0)}\Bigr)}\right); 
\end{equation} 
see, e.g., the classical book by Risken~\cite{1989fpem.book.....R}.
Therefore, one gains any multi-time 
correlation functions through appropriate integration with~\eqref{sol_prob_exact_massive} in the free massive case. 
By definition and with the use of Markov's property, the two-time correlation function, $t_1 \geq t_2\geq t_0$, is
\begin{align}
\Bigl\langle \varphi (t_1) \,  \varphi (t_2) \Bigl| \bigl[ \varphi_0, t_0\bigr] \Bigr \rangle & = \int\limits_{-\infty}^{+\infty} \int\limits_{-\infty}^{+\infty} d 
{\varphi}_1 \,  d 
{\varphi}_2 \, 
{\varphi}_1 \, 
{\varphi}_2 \,  \mathcal{P}_m \bigl( \, 
{\varphi}_1, t_1 ; \, 
{\varphi}_2, t_2 \, \bigl| \,
{\varphi}_0 , t_0 
\bigr) \nonumber \\ \label{two-point_t1t2}
& \,\, = \int\limits_{-\infty}^{+\infty} \int\limits_{-\infty}^{+\infty} d 
{\varphi}_1 \,  d 
{\varphi}_2 \, 
{\varphi}_1 \, 
{\varphi}_2 \, \mathcal{P}_m \bigl( \, 
{\varphi}_1, t_1 \, \bigl| \, 
{\varphi}_2, t_2 \bigr) \, \mathcal{P}_m \bigl( 
{\varphi}_2 , t_2 \, \bigl| \, \varphi_0, t_0 
\bigr). \end{align} 
Without any assumption on initial values of $\varphi_0$ and $t_0$, through~\eqref{two-point_t1t2} with~\eqref{sol_prob_exact_massive}, one has 
\begin{align}\label{O-U_correlator-p0t0}
\Bigl\langle \varphi (t_1) \,  \varphi (t_2) \Bigl| \bigl[ \varphi_0, t_0\bigr] 
\Bigr \rangle 
& = \frac{3H^4}{8 \, \pi^2m^2} \, \biggl( \e^{-\tfrac{m^2}{3H}(t_1-t_2)} - \e^{-\tfrac{m^2}{3H}(t_1+t_2-2t_0)} \biggr) \\ \nonumber
& \qquad + \, \frac{ \, \varphi_0^2 \, \biggl( \e^{-\tfrac{m^2}{3H}(t_1+t_2-2t_0)} - \e^{-\tfrac{m^2}{3H}(t_1+3t_2-4t_0)} \biggr)}{\Bigl(1-\e^{-\tfrac{2m^2}{3H}(t_2-t_0)}\Bigr)}.  
\end{align} We set up the initial value~$\varphi_0=0$ at time $t_0=0$ such that $ \left. \bigl\langle \varphi (t_0) \varphi (t_0) \bigl| \bigl[ \varphi_0, t_0\bigr]\bigr \rangle \right|_{t_0=0} =0$. The chosen values $\varphi_0=t_0=0$  lead to the following form of the conditional (or transition) probability $\mathcal{P}_m \bigl( \varphi_2, t_2 \bigl| 0, 0 \bigr)$ in~\eqref{sol_prob_exact_massive}:
\begin{equation}\label{Prob_x2t2}
 \mathcal{P}_m \bigl( \varphi_2 , t_2 \bigr) = \,\, \frac{2\sqrt{\pi} \, m}{\sqrt{3} \, H^2 \,
} \biggl(1-\e^{-\tfrac{2m^2t_2}{3H}}\biggr)^{-1/2} 
\text{exp} \, \left(-\frac{4\pi^2m^2 \varphi_2^2}{3H^4\biggl(1-\e^{-\tfrac{2m^2t_2}{3H}}\biggr)} \right).
\end{equation}
Remarkably, the transition probability $\mathcal{P}_m \bigl( \varphi_2, t_2 \bigl| \, \varphi_0, t_0 \bigr)$ at the limit $t_0 \rightarrow - \infty$ goes to the stationary distribution:
\begin{equation}\label{limit_to_inf}
\mathcal{P}_m \bigl( \varphi_2, t_2 \bigl| \, \varphi_0, t_0 \bigr) \xrightarrow[\,\,\,\,]{\,\, t_0 \rightarrow \, - \,  \infty \,\,} 
\mathcal{P}_{\text{st}} \bigl( \varphi_2 \bigr) :=\frac{2\sqrt{\pi} \, m}{\sqrt{3} \, H^2 \,} \,\,
\text{exp} \, \left( \, -\frac{4\pi^2m^2 \varphi_2^2}{3H^4} \right),   
\end{equation} and the two-time correlation function~\eqref{O-U_correlator-p0t0} depends only on the time difference: 
\begin{equation}\label{O-U_correlator_late}
\left. \Bigl\langle \varphi (t_1) \,  \varphi (t_2) \Bigl| \bigl[ \varphi_0, t_0\bigr] 
\Bigr \rangle \right|_{t_1 \geq t_2} \xrightarrow[]{\,\, t_0 \rightarrow \, - \infty \,\,} \frac{3H^4}{8 \, \pi^2m^2} \,  \e^{-\tfrac{m^2}{3H}(t_1-t_2)}.  
\end{equation} 
That is the notable feature of the stationary process, to which belong the Fokker-Planck or forward Kolmogorov equation~\eqref{Fokker-Planck_General_t1t2} with the linear drift term, i.e., in the free massive case without self-interaction. By inserting the chosen initial values $\varphi_0=t_0=0$ into~\eqref{O-U_correlator-p0t0} or, equivalently, after straightforward integration of~\eqref{two-point_t1t2} with~\eqref{sol_prob_exact_massive} and~\eqref{Prob_x2t2}, one gets 
\begin{equation}\label{O-U_correlator}
\Bigl\langle \varphi (t_1) \,  \varphi (t_2) \Bigl| \bigl[ 0, 0\bigr] 
\Bigr \rangle:=\Bigl\langle \varphi (t_1) \,  \varphi (t_2) 
\Bigr \rangle = \frac{3H^4}{8 \, \pi^2m^2} \, \biggl( \e^{-\tfrac{m^2}{3H}(t_1-t_2)} - \e^{-\tfrac{m^2}{3H}(t_1+t_2)} \biggr) .  
\end{equation} 
This is the well-known correlation function of the Ornstein-Uhlenbeck mean-reverting stochastic process~\cite{Uhlenbeck:1930zz,1989fpem.book.....R}: the unique Gaussian and Markov process that has the stationary state. The obtained two-point correlation function~\eqref{O-U_correlator} precisely coincides with the free massive scalar field’s one~\eqref{O-U_correlator_QFT} from our previous work~\cite{2025JHEP...04..127K} and from~\cite{2014PhRvD..89f3506G,2015PhRvD..91f3520G}. By its virtue,~\eqref{O-U_correlator} tends to the equilibrium state, which depends only on the time difference and turns out to be de Sitter-invariant. That attractor feature in de Sitter space was also revealed in the preceding works~\cite{1994PhRvD..50.6357S,2000PhRvD..62l4019A} from different perspectives. One can notice that at equal times~\eqref{O-U_correlator} matches~\eqref{2pt_massive_free} above, which can be obtained with the help of~\eqref{Prob_x2t2} via the straightforward integration:
\begin{equation}\label{two-point_tt}
\Bigl\langle \varphi^2 (t) \Bigl| \bigl[ 0, 0\bigr] \Bigr \rangle := \Bigl\langle \varphi^2 (t) \Bigr \rangle = \int\limits_{-\infty}^{+\infty} d 
{\varphi} \, \,  
{\varphi}^2 \, \mathcal{P}_m \bigl( 
{\varphi} , t \bigr) = \frac{3H^4}{8 \, \pi^2m^2} \, \biggl( 1 - \e^{-\tfrac{2m^2t}{3H}} \biggr). \end{equation} 
Indeed,~\eqref{Prob_x2t2} is exactly our ansatz~\eqref{ansatz_massive} with~\eqref{eq_for_ansatz} from above. Let us point out that this two-point function not only reduces to Wiener process’s one~\cite{Vilenkin:1982wt,Linde:1982uu} in the smooth massless limit but also mimics this behavior in the regime $ t \ll H/m^2$, as it was observed in~\cite{1983PhRvD..27.2848V}.

To get the iterative $\lambda$ series for the multi-time correlation functions, one employs~\eqref{Fokker-Planck_General_t1t2} with replaced $t$ with $t_1$ and $\varphi$ with $\varphi_1$. In the two-time correlation function's case, we multiply l.h.s. 
and r.h.s. of~\eqref{Fokker-Planck_General_t1t2} by $\varphi_1$, $\varphi_2$, and $\mathcal{P}_m \bigl( \varphi_2 , t_2 \bigr)$, integrate over $\varphi_1$, $\varphi_2$, and gain the corresponding first-order differential equation. 
Its formal solution can be written~as 
\begin{equation}\label{2pt_t1t2_gen_integral}
\Bigl\langle \varphi (t_1) \,  \varphi (t_2) \Bigr \rangle_{\kk+1} = - \frac{\lambda}{3H} \, \e^{-\tfrac{m^2t_1}{3H}} \int dt_1 \, \e^{\tfrac{m^2t_1}{3H}} \,  \Bigl\langle \varphi^3 (t_1) \,  \varphi (t_2) \Bigr \rangle_{\kk}. 
\end{equation} In contrast to the equal-time correlation functions, we have here the indefinite integral. After the integration, one has the antiderivative and the unknown function, which one defines by matching it to the known result for the equal-time correlation function. 
In the most simple example, namely, for the $\lambda${--}linear order in~\eqref{2pt_t1t2_gen_integral}, we will illustrate the proposed technique in detail. The integrand we extract through the definition with the use of the Markov property and with the initial conditions $\varphi_0=t_0=0$ as above:
\begin{align}\label{4-pt_3t1t2_1}
\Bigl\langle \varphi^3 (t_1) \,  \varphi (t_2) \Bigr \rangle_0 & = \int\limits_{-\infty}^{+\infty} \int\limits_{-\infty}^{+\infty} d 
{\varphi}_1 \,  d 
{\varphi}_2 \, \, 
{\varphi}^3_1 \, 
{\varphi}_2 \, \, \mathcal{P}_m \bigl( \, 
{\varphi}_1, t_1 \, \bigl| \, 
{\varphi}_2, t_2 \bigr) \, \mathcal{P}_m \bigl( 
{\varphi}_2 , t_2 \bigr) \\ \label{4-pt_3t1t2_2} & = \frac{27H^8}{64 \, \pi^4 m^4}  \biggl( 1 - \e^{-\tfrac{2m^2 t_1}{3H}} \biggr)  \biggl( \e^{-\tfrac{m^2}{3H}(t_1-t_2)} - \e^{-\tfrac{m^2}{3H}(t_1+t_2)} \biggr). 
\end{align}
Therefore, one calculates indefinite integral~\eqref{2pt_t1t2_gen_integral} with the obtained integrand~\eqref{4-pt_3t1t2_2}:
\begin{equation}\label{2pt_t1t2_C(t2)}
\Bigl\langle \varphi (t_1) \,  \varphi (t_2) \Bigr \rangle_\lambda = - \frac{27 \lambda H^8}{64\, \pi^4 m^6} \Biggl( \biggl(\e^{\tfrac{m^2 t_2}{3H}} - \e^{-\tfrac{m^2t_2}{3H}} \biggr) \biggl( \frac{m^2 t_1}{3H} + \frac{1}{2} \,  \e^{-\tfrac{2m^2 t_1}{3H}}\biggr) + 
\mathrm{C} \bigl(t_2 \bigr) \Biggr) \e^{-\tfrac{m^2 t_1}{3H}}.
\end{equation} 
Here, $\mathrm{C} \, (t_2)$ is the unknown function of the only argument $t_2$. 
To define it, one can rely on the known answer for $\bigl\langle \varphi^2 (t) \bigr \rangle_\lambda$ from~\eqref{2pt_massive_lambda_series}. We put $t_1 = t_2$ in this expression above~\eqref{2pt_t1t2_C(t2)} and equate it to~\eqref{2pt_massive_lambda_series}, where we substitute $t= t_2$ in the $\lambda${--}linear order, resulting in
\begin{align}\label{2pt_t1t2_full_linearlambda}
\Bigl\langle \varphi (t_1) \,  \varphi (t_2) \Bigr \rangle_\lambda  = & - \frac{27 \lambda H^8}{128 \, \pi^4 m^6}\Biggl( \biggl(2+ \frac{2m^2}{3H} \bigl(t_1-t_2\bigr) \biggr)\e^{-\tfrac{m^2}{3H}(t_1-t_2)} + \, \e^{-\tfrac{m^2}{3H}(3 \, t_1-t_2)} \\ \nonumber 
& - \biggl(1 + \frac{2m^2}{3H} \bigl(t_1 + 3 t_2\bigr) \biggr)\e^{-\tfrac{m^2}{3H}(t_1+t_2)} 
- \e^{-\tfrac{m^2}{3H}(3 \,t_1+t_2)} - \e^{-\tfrac{m^2}{3H}(t_1 + 3 \, t_2)}\Biggr).
\end{align} 

One can proceed in this way to calculate the four-point correlation function. The idea is the same: we relate different $2\n$-point correlation functions at the different $\lambda$ orders through the Fokker-Planck or the forward Kolmogorov equation~\eqref{Fokker-Planck_General_t1t2}. We intend to compute the four-point correlation function at the linear $\lambda$ order, i.e., $\bigl\langle \varphi (t_1) \,  \varphi (t_2) \,  \varphi (t_3) \,  \varphi (t_4)\bigr \rangle_\lambda$. Not surprising, to find that one needs to have at hand the six-point correlation function at the zero order and the two-point one at the $\lambda${--}linear one. At the zero $\lambda$ order, one calculates the integrand through the definition with the same initial conditions $\varphi_0=t_0=0$ and using the Markov property, $t_1\geq t_2\geq t_3 \geq t_4\geq t_0$:
\begin{align}\label{integrand_four-point_t1t2}
& \Bigl\langle \varphi^3 (t_1) \,  \varphi (t_2) \,  \varphi (t_3) \,  \varphi (t_4)\Bigr \rangle_0 = \\ & \qquad \qquad \qquad = \int\limits_{-\infty}^{+\infty}.. \int\limits_{-\infty}^{+\infty} 
\biggl( \, \,  \prod\limits_{i=1}^4 \, d
{\varphi}_i \biggr)
\, \,  
{\varphi}^3_1 \, 
{\varphi}_2 \, 
{\varphi}_3 \, 
{\varphi}_4 \,  \biggl( \,\,\prod\limits_{i=1}^3  \mathcal{P}_m \bigl( \, 
{\varphi}_i, t_i \, \bigl| \, 
{\varphi}_{i+1}, t_{i+1} \bigr) \biggr) \, \mathcal{P}_m \bigl( 
{\varphi}_4 , t_4 \bigr) \nonumber \\ \label{integrand_four-point_t1t2_answer}
& \qquad \qquad \qquad 
= \frac{81H^{12}}{512 \, \pi^6 m^6} \, \e^{-\tfrac{m^2t_1}{3H}} \, \biggl( 1 - \e^{-\tfrac{2m^2 t_4}{3H}} \biggr)  \Biggl( \e^{-\tfrac{2m^2 t_1}{3H}} \biggl(5\, \e^{-\tfrac{m^2}{3H}(t_2+t_3-t_4)} \\ \nonumber  & \qquad \qquad \qquad \qquad \qquad \qquad \,\,\, 
- 4\, \e^{-\tfrac{m^2}{3H}(t_2-t_3-t_4)} - 3\, \e^{\,\, \tfrac{m^2}{3H}(t_2-t_3+t_4)} + 2 \, \e^{\,\, \tfrac{m^2}{3H}(t_2+t_3+t_4)} \biggr) \\ \nonumber  & \qquad \qquad \qquad \qquad \qquad \qquad \,\,\, 
- \, 3 \, \e^{-\tfrac{m^2}{3H}(t_2+t_3-t_4)} + \, 2 \, \e^{-\tfrac{m^2}{3H}(t_2-t_3-t_4)} +  \e^{\, \, \tfrac{m^2}{3H}(t_2-t_3+t_4)} \Biggr).
\end{align} We have used the integral expressions from appendix~\ref{appendix}. By setting $t_1=t_2=t_3=t_4 := t$ in~\eqref{integrand_four-point_t1t2_answer}, one comes back to~\eqref{2pt_massive_free}, as expected.

We shall start with the case where the last time moment differs, while other three coincide. 
By multiplying both l.h.s.~\eqref{Fokker-Planck_General_t1t2} 
and r.h.s.~\eqref{Fokker-Planck_General_t1t2} (with replaced  $t$ with $t_1$ and $\varphi$ with $\varphi_1$) by $\varphi^3_1$, $\varphi_2$, $\mathcal{P}_m \bigl( \varphi_2 , t_2 \bigr)$, integrating over $\varphi_1$, $\varphi_2$, and denoting $t_2:= t_4$, one arrives at the corresponding first-order differential equation, whose formal solution takes the form:
\begin{equation}\label{4pt_3t1_t4_def}
\Bigl\langle \varphi^3 (t_1) \,  \varphi (t_4) \Bigr \rangle_{\kk+1} 
= \e^{-\tfrac{m^2 t_1}{H}} \int \, dt_1 \, \e^{\tfrac{m^2 t_1}{H}} \Biggl(- \frac{\lambda}{H} \, \Bigl\langle \varphi^{5} (t_1) \, \varphi (t_4) \Bigr \rangle_{\kk} 
+ \, \frac{3 H^3}{4\pi^2} \, \Bigl\langle \varphi (t_1) \,  \varphi (t_4) \Bigr \rangle_{\kk+1} 
\, \Biggr).
\end{equation}
For the computation of the $\lambda${--}linear order there, we already know the integrand from~\eqref{2pt_t1t2_full_linearlambda} and~\eqref{integrand_four-point_t1t2_answer}. 
By taking indefinite integral~\eqref{4pt_3t1_t4_def}, one gets an expression with
the unknown function $\mathrm{C}\, (t_4)$ and defines it as in the previous case: by substituting $t_1=t_4$ in the indefinite integral' expression 
and equating it to $\bigl\langle \varphi^4 (t_4) \bigr\rangle_\lambda$ from~\eqref{4pt_massive_lambda_series}. 
Following the same spirit, one finds the corresponding equation for another four-point correlation function. Once again, we multiply 
both l.h.s.~\eqref{Fokker-Planck_General_t1t2} and r.h.s.~\eqref{Fokker-Planck_General_t1t2}, with replaced $t$ with $t_1$, $\varphi$ with $\varphi_1$, and $t_0$ with $t_3$ here, by $\varphi^2_1$, $\varphi_3$, $\varphi_4$, $\mathcal{P}_m \bigl( \varphi_3, t_3 \, \bigl| \, \varphi_4, t_4 \bigr)$, $\mathcal{P}_m \bigl( \varphi_4 , t_4 \bigr)$ and integrate over $\varphi_1$, $\varphi_3$ and~$\varphi_4$. The formal solution in this case becomes
\begin{align}\label{4pt_2t1_t3_t4_def}
\Bigl\langle \varphi^2 (t_1) \, \varphi (t_3) \,  \varphi (t_4) \Bigr \rangle_{\kk+1} 
= \, \e^{-\tfrac{2m^2 t_1}{3H}} \int \, dt_1 \, \e^{\tfrac{2m^2 t_1}{3H}} \Biggl( & - \frac{ 2\lambda}{3H} \, \Bigl\langle \varphi^{4} (t_1) \, \varphi (t_3) \, \varphi (t_4) \Bigr \rangle_{\kk} \\ \nonumber 
& \quad + \, \frac{H^3}{4\pi^2} \, \Bigl\langle \varphi (t_3) \,  \varphi (t_4) \Bigr \rangle_{\kk+1} 
\, \Biggr).
\end{align} 
For the $\lambda${--}linear order, after straightforward integration one has the indefinite integral and
the unknown function $\mathrm{C}\, \bigl(t_3, t_4\bigr)$, which we define by matching it to the previous "iteration" $\bigl\langle \varphi^3 (t_1) \varphi (t_4) \bigr \rangle_\lambda$. 
As the final step, one 
gets the analogous equation, and
its formal solution can be written in the same manner as above:
\begin{equation}\label{four-point_diff_time_formal_sol}
\Bigl\langle \varphi (t_1) \, \varphi (t_2) \, \varphi (t_3) \, \varphi (t_4) \Bigr \rangle_{\kk+1}
= - \frac{\lambda}{3H} \, \e^{-\tfrac{m^2t_1}{3H}} \int dt_1 \, \e^{\tfrac{m^2 t_1}{3H}}  \Bigl\langle \varphi^3 (t_1) \varphi (t_2) \varphi (t_3) \varphi (t_4) \Bigr \rangle_{\kk}.
\end{equation}
For the $\lambda${--}linear order, that integral appears to be some expression with
the unknown function $\mathrm{C}\, \bigl(t_2, t_3, t_4\bigr)$, defined by matching  
to 
$\bigl\langle \varphi^2 (t_1)  \,  \varphi (t_3) \,  \varphi (t_4) \bigr \rangle_\lambda$ and setting $t_2 = t_1$. 
The final result after all these steps is 
\begin{align}
\label{four_point_OneLoop}
& \bigl\langle \varphi \left(t_1\right) \varphi \left(t_2\right)  \varphi \left(t_3\right) \varphi \left(t_4\right) \bigr\rangle_{\lambda} = - \frac{81 \lambda H^{12}}{1024 \, \pi^6m^8} \Biggl( \Bigl(4 + \frac{2m^2}{3H} \bigl(t_1-t_2+t_3-t_4 \bigr)\Bigr) \e^{-\tfrac{m^2}{3H}\left(t_1-t_2+t_3-t_4\right)} \nonumber \\
& 
\,\,\, + \Bigl(16 + \frac{4m^2}{3H} \bigl(t_1+3\, t_2-3\, t_3-t_4 \bigr)\Bigr) \, \e^{-\tfrac{m^2}{3H}\left(t_1+t_2-t_3-t_4\right)} - 2 \, \e^{-\tfrac{m^2}{3H}(t_1+t_2+t_3- 3\, t_4)} \\ \nonumber 
& 
\,\,\, - 2 \, \e^{-\tfrac{m^2}{3H}\left(3t_1 - t_2 - t_3 - t_4\right)} - \Bigl( 3 + \frac{2m^2}{3H} \bigl(t_1 - t_2 + t_3+ 3\, t_4 \bigr)\Bigr) \, \e^{-\tfrac{m^2}{3H}(t_1-t_2+t_3+t_4)} \\ \nonumber 
& 
\,\,\, - \Bigl(14 + \frac{4m^2}{3H} \bigl(t_1+3\, t_2-3\, t_3+3\, t_4 \bigr)\Bigr) \, \e^{-\tfrac{m^2}{3H}(t_1+t_2-t_3+t_4)} + \e^{-\tfrac{m^2}{3H}\left(t_1-t_2+3t_3-t_4\right)} \\ \nonumber 
& 
\,\,\, - \Bigl( 33 + \frac{2m^2}{H} \bigl(t_1+3\,t_2+5 \,t_3-5\, t_4 \bigr)\Bigr) \, \e^{-\tfrac{m^2}{3H}(t_1+t_2+t_3-t_4)} + 2 \, \e^{-\tfrac{m^2}{3H}\left(3t_1-t_2-t_3+t_4\right)} \\ \nonumber
& 
\,\,\, + 3 \, \e^{-\tfrac{m^2}{3H}\left(3t_1-t_2+t_3-t_4\right)} + 4 \, \e^{-\tfrac{m^2}{3H}(3t_1+t_2-t_3-t_4)} + 4 \, \e^{-\tfrac{m^2}{3H}(t_1+3t_2-t_3- t_4)} - \e^{-\tfrac{m^2}{3H}(t_1-t_2+3t_3+t_4)} \\ \nonumber 
& 
\,\,\, - \e^{-\tfrac{m^2}{3H}(t_1-t_2+t_3+3t_4)} + \Bigl(30 + \frac{2m^2}{H}\bigl(t_1+3 \, t_2+5\, t_3+7\, t_4\bigr) \Bigr)\e^{-\tfrac{m^2}{3H}(t_1+t_2+t_3+t_4)} \\ \nonumber 
& 
\,\,\,  - 2 \, \e^{-\tfrac{m^2}{3H}(t_1+t_2-t_3+3t_4)} -  3 \, \e^{-\tfrac{m^2}{3H}(3t_1-t_2+t_3+t_4)} - 4 \, \e^{-\tfrac{m^2}{3H}(3t_1+t_2-t_3+t_4)} - 5 \, \e^{-\tfrac{m^2}{3H}(3t_1+t_2+t_3-t_4)} \\ \nonumber
& 
\,\,\, - 4 \, \e^{-\tfrac{m^2}{3H}(t_1+3t_2-t_3+t_4)} - 5 \, \e^{-\tfrac{m^2}{3H}(t_1+3t_2+t_3-t_4)} - 5 \, \e^{-\tfrac{m^2}{3H}(t_1+t_2+3t_3-t_4)} + 5 \, \e^{-\tfrac{m^2}{3H}(3t_1+t_2+t_3+t_4)} \\ \nonumber 
& 
\,\,\,  + 5 \, \e^{-\tfrac{m^2}{3H}(t_1+3t_2+t_3+t_4)} + 5 \, \e^{-\tfrac{m^2}{3H}(t_1+t_2+3t_3+t_4)} + 5 \, \e^{-\tfrac{m^2}{3H}(t_1+t_2+t_3+3t_4)} \Biggr).
\end{align} 
The last comparison will be made for the two-point correlation function at the $\lambda^2$ order. Our equation is~\eqref{2pt_t1t2_gen_integral} 
and $ \bigl\langle \varphi^3 (t_1) \,  \varphi (t_2) \bigr \rangle_\lambda$ is taken from~\eqref{four_point_OneLoop}. Following our proposed routine, one takes the indefinite integral 
and finds the unknown function $\mathrm{C} (t_2)$ by matching this expression to $\bigl\langle\varphi^2(t_2) \bigr \rangle_{\lambda^2}$ from~\eqref{2pt_massive_lambda_series}, 
resulting in
\begin{align}\label{2pt_t1t2_squared_lambda_full_answer}
& \Bigl\langle \varphi (t_1) \,  \varphi (t_2) \Bigr \rangle_{\lambda^2}  = \frac{81 \, \lambda^2 H^{12}}{2048 \, \pi^6 m^{10}} \Biggl( \biggl( 30 + \frac{12 m^2}{H} \, \bigl(t_1-t_2\bigr) + \frac{2m^4}{3H^2} \, \bigl(t_1-t_2\bigr)^2 \biggr) \, \e^{-\tfrac{m^2}{3H}(t_1-t_2)} \nonumber \\
&\qquad \quad + 2 \, \e^{-\tfrac{3m^2}{3H}(t_1-t_2)} + \biggl( 48 + \frac{2 m^2}{H} \, \bigl(9\, t_1 - 5\, t_2\bigr) \biggr) \, \e^{-\tfrac{m^2}{3H}(3t_1-t_2)} \\ 
\nonumber 
& \qquad \quad + \biggl( 36 - \frac{2 m^2}{H} \, \bigl( 5\, t_1 + 23\, t_2 \bigr) - \frac{2m^4}{3H^2} \,\bigl( t_1 + 3\, t_2\bigr)^2 \biggr) \, \e^{-\tfrac{m^2}{3H}(t_1+t_2)} \\ \nonumber 
& \qquad \quad - \biggl( 45 + \frac{2 m^2}{H} \, \bigl( 9\, t_1 + 7\,t_2\bigr) \biggr) \, \e^{-\tfrac{m^2}{3H}(3t_1+t_2)} -   \biggl( \frac{117}{2} + \frac{2 m^2}{H} \, \bigl( t_1 + 15 \,t_2\bigr) \biggr) \, \e^{-\tfrac{m^2}{3H}(t_1+3t_2)} \\  \nonumber 
& \qquad \quad  + \frac{15}{2} \, \e^{-\tfrac{m^2}{3H}(5t_1-t_2)} - \frac{15}{2} \, \e^{-\tfrac{m^2}{3H}(5t_1+t_2)} - 5 \, \e^{-\tfrac{m^2}{3H}(3t_1+3t_2)}  - \frac{15}{2} \, \e^{-\tfrac{m^2}{3H}(t_1+5t_2)}\Biggr). 
\end{align}  
All the obtained expressions from this section coincide with ones from our previous work~\cite{2025JHEP...04..127K}, converge, and at the late-time limit tend to the de Sitter-invariant ones; see~\eqref{loop_series_late_time} below.
The equations preceding the formal solutions~\eqref{2pt_t1t2_gen_integral} and~\eqref{4pt_3t1_t4_def}{--}\eqref{four-point_diff_time_formal_sol} in the free massive case, i.e., $\lambda=0$, are similar to the generalized "formulae of differentiation" from the original work by Shapiro and Loginov~\cite{SHAPIRO1978563} for the Gaussian process; see also references~\cite{1992sppc.book.....V,etde_634926}.

For the de Sitter-invariant two-time correlation functions, through our approach here, one can gain perturbative series as well. 
One takes the late-time behavior of $\bigl\langle \varphi^3 (t_1) \,  \varphi (t_2) \bigr\rangle_0$ from~\eqref{4-pt_3t1t2_2} and $\bigl\langle \varphi^3 (t_1) \,  \varphi (t_2) \bigr\rangle_\lambda$ from~\eqref{four_point_OneLoop}, which depends on time difference, integrates through \eqref{2pt_t1t2_gen_integral}, and matches it to equal-time results at late times, which are just a constant:
\begin{align}
\Bigl\langle \varphi^3 (t_1) \,  \varphi (t_2) \Bigr \rangle_0 &\xrightarrow[\ \text{times}]{\text{late}} \,\, \frac{27 \, H^8}{64 \, \pi^4 m^4} \,\, \e^{-\tfrac{m^2}{3H}(t_1-t_2)} \, \quad \Rightarrow \\ \nonumber 
\Bigl\langle \varphi (t_1) \, \varphi (t_2) \Bigr \rangle_\lambda \,\, & \stackrel{ \eqref{2pt_t1t2_gen_integral}}{=} \left. - \frac{9 \, \lambda H^7}{64 \pi^4 m^4} \Bigl( t_1 + \mathrm{C} \bigl(t_2\bigr) \Bigr) \e^{-\tfrac{m^2}{3H}(t_1-t_2)} \right|_{t_1=t_2} 
= \bigl\langle \varphi^2 \bigr \rangle^{\text{late-time}}_\lambda \stackrel{ \eqref{two-point_equilibrium}}{=} - \frac{27 \, \lambda H^8}{64 \pi^4 m^6}. 
\end{align} Therefore, the final de Sitter-invariant series is the following:
\begin{align} \label{loop_series_late_time} 
& \Bigl\langle \varphi (t_1) \,  \varphi (t_2) \Bigr \rangle \xrightarrow[\, \text{times}\,]{\text{late}}  
\frac{3H^4}{8\pi^2m^2} \, \e^{-\tfrac{m^2}{3H}\left(t_1-t_2\right)} - \frac{27 \, \lambda H^8}{64 \pi^4m^6} \biggl(1+\frac{m^2}{3H}\left(t_1 -t_2 \right) \biggr) \e^{-\tfrac{m^2}{3H}\left(t_1-t_2\right)} \\ \nonumber  
& 
\, + \frac{81\, \lambda^2 H^{12}}{1024 \, \pi^6 m^{10}} \biggl(15+\frac{6m^2}{H}\left(t_1 -t_2 \right) + \frac{m^4}{3H^2}\left(t_1 - t_2 \right)^2 + \e^{-\tfrac{2m^2}{3H}\left(t_1 - t_2 \right)} \biggr)\e^{-\tfrac{m^2}{3H}\left(t_1-t_2\right)} + O\left(\lambda^3\right). \end{align} 
The obtained series is in agreement with the direct computations within the Schwinger-Keldysh formalism~\cite{2025JHEP...04..127K,2022EPJC...82..345K,2013PhLB..727..541G} and corresponds to the late-time limit of~\eqref{O-U_correlator},~\eqref{2pt_t1t2_full_linearlambda}, and~\eqref{2pt_t1t2_squared_lambda_full_answer}. In our previous work~\cite{2025JHEP...04..127K}, we have established the correspondence between integral structures arising in the Yang-Feldman-type equation and in the Feynman or the Schwinger-Keldysh diagrammatic techniques. In the equilibrium, authors of~\cite{2014PhRvD..89f3506G} extracted the topologically distinct diagrams for the equal-time two-point correlation function. They employed both the stochastic approach and the closed-time-path formalism. Contributions for each type of diagram in~\cite{2025JHEP...04..127K} and the series~\eqref{two-point_equilibrium} up to the $\lambda^2$ at the equilibrium match those derived in~\cite{2014PhRvD..89f3506G}. We also point out that the types of integral structures arising within the Yang-Feldman formalism~\cite{2025JHEP...04..127K} correspond to the diagrammatic stochastic rules in~\cite{2014PhRvD..89f3506G} with an appropriate assignment in between. Remarkably, our proposed technique in the present work takes several trivial lines, in contrast to alternative approaches.

We conclude this section with the massless case. The exact solution to~\eqref{Fokker-Planck_General_t1t2} for the conditional (or transition) probability is known:
\begin{equation}\label{sol_prob_exact_massless}
\mathcal{P}_{m=0} \bigl( \varphi, t \, \bigl| \, \varphi_0, t_0 \bigr) = \frac{\sqrt{2\pi} }{\sqrt{\,H^3(t-t_0) \,}} \,\, \text{exp}\left(- \frac{2\pi^2 \, \bigl( \varphi \, - \, \varphi_0 \bigr)^2}{\,H^3(t-t_0) \,}\right). 
\end{equation} One finds the analogue to~\eqref{Prob_x2t2} 
and the correlation function by straightforward integration as before in~\eqref{two-point_t1t2},~\eqref{4-pt_3t1t2_1},~\eqref{integrand_four-point_t1t2}, etc. With arbitrary $\varphi_0$ and $t_0$ values, through~\eqref{two-point_t1t2} we~have 
\begin{equation}
\label{O-U_correlator_masless}
\left. \Bigl\langle \varphi_{m=0} (t_1) \,  \varphi_{m=0} (t_2) \Bigl| \bigl[ \varphi_0, t_0\bigr] 
\Bigr \rangle \right|_{t_1 \geq t_2}
= \frac{H^3}{4\pi^2} \bigl(t_2-t_0\bigr) + \varphi_0^2,   
\end{equation} being the smooth massless limit of~\eqref{O-U_correlator-p0t0}. To grant $ \left. \bigl\langle \varphi_{m=0} (t_0) \, \varphi_{m=0} (t_0) \bigl| \bigl[ \varphi_0, t_0\bigr]  \bigr \rangle \right|_{t_0 =0} =0$, we set~up $\varphi_0=0$ at the initial time moment $t_0=0$ as before. Following our approach, one defines the necessary first-order differential equation and, from its formal solution (without exponents), takes the indefinite integral. The result of integration should match equal-time (or the corresponding previous "iteration") ones to fix the unknown function. We present the final expressions with the chosen initial conditions $\varphi_0=t_0=0$ for the massless case: 
\begin{align}
& \qquad \qquad \qquad \qquad \,\,\, \Bigl\langle \varphi_{m=0} (t_1) \,  \varphi_{m=0} (t_2) \Bigr \rangle = \frac{H^3t_2}{4\pi^2} - \frac{\lambda H^5}{96 \, \pi^4} \Bigl( 3\, t^2_1t_2 + t^3_2 \Bigr) \\ \nonumber 
& \qquad \qquad \qquad \qquad \qquad \qquad \qquad \qquad \qquad \quad 
+  \frac{\lambda^2 H^7}{1536 \, \pi^6} \biggl( 11\, t^4_1 t_2 + 2\, t_1^2 t^3_2+ \frac{31}{5} \, t^5_2 \biggr) + O\bigl(\lambda^3\bigr) ; 
\\ 
& \Bigl\langle \varphi^3_{m=0} (t_1) \,  \varphi_{m=0} (t_2) \,  \varphi_{m=0} (t_3) \,  \varphi_{m=0} (t_4) \Bigr \rangle_0 = \frac{3H^9}{64 \, \pi^6} \, \Bigl(t_1t_2t_4 + 2\, t_1 t_3 t_4 + 2\, t_2 t_3 t_4 \Bigr); \\
&\Bigl\langle \varphi_{m=0} (t_1) \,  \varphi_{m=0} (t_2) \,  \varphi_{m=0} (t_3) \,  \varphi_{m=0} (t_4) \Bigr \rangle_\lambda  = - \frac{\lambda H^8}{384 \, \pi^6} \, \Bigl(3 \, t_1^2 \, t_2 t_4 +6 \, t_1^2 \, t_3 t_4 \\ \nonumber 
& \qquad \qquad \qquad \qquad \qquad \quad + 12 \, t_1 t_2 t_3 t_4 + t_2^3 \, t_4 + 6 \, t_2^2\,  t_3 t_4 
+ \, 3 \, t_2t_3^2 \,t_4 + t_2t_4^3 + 2 \, t_3^3 \,t_4 + 2t_3t_4^3 \Bigr). 
\end{align}
Certainly, all results listed above are the massless limit of~\eqref{O-U_correlator}
,~\eqref{2pt_t1t2_full_linearlambda},~\eqref{integrand_four-point_t1t2_answer},~\eqref{four_point_OneLoop}, and~\eqref{2pt_t1t2_squared_lambda_full_answer}, and two-point and four-point correlation functions match those obtained in~\cite{2025JHEP...04..127K}. At the equal times, those expressions shrink to~\eqref{2pt_massless_series}{--}\eqref{6pt_massless_series}. 

\section{Conclusion}\label{Conclusion}
We proposed an alternative approach to gain the equal-time and multi-time correlation functions for the spectator field in de Sitter space directly through the Fokker-Planck or forward Kolmogorov equation. For a rather particular theory of a minimally coupled scalar field~\eqref{S0}, we derived the first-order differential equation through the Fokker-Planck equation. Its solution in the integral form relates various correlation functions at the different self-interaction coupling constant $\lambda$ orders. One solves it iteratively. The starting point is the setup without self-interaction, and the exact solutions are well-known there. Further, we engage these correlation functions~to ones at the different $\lambda$ orders by the integral relation. In the case of equal-time correlation functions, that relation provides the results order by order in $\lambda$ through the straightforward integration~\eqref{reccurent_equation_massive_and_lambda}. In the multi-time case, we have the formal solution to the corresponding first-order differential equation in the indefinite integral' form. After the integration, one has the antiderivative and the unknown function, which one defines by matching it to the known result for the equal-time correlation function step by step; see section~\ref{Section_multi}. The obtained results are in agreement with the corresponding quantum field theory's perturbative series for two-point and four-point correlation functions. Remarkably, the massive perturbative on the $\lambda$ series converges and fully resumms the well-known massless time divergences of the self-interacting scalar field in de Sitter space, while at the smooth massless limit the massive series reduces to the known massless results. This work provides additional confirmation that Starobinsky's stochastic approach reproduces precisely quantum field theory's results on de Sitter space for scalar field models.

\acknowledgments We are grateful to the anonymous referee for the proposed improvements to the results presentation. P.P. would like to thank David Brizuela for the kind hospitality at the University of the Basque Country. 

\appendix 
\section{{Where is the leading infrared logarithm in the massive series?}}\label{Leading_log}
In this Appendix we present some details to endorse the statement regarding the expansion scheme in section~\ref{Section_equal}. We consider as an example the two-point correlation function~\eqref{2pt_massive_lambda_series}. By expanding the obtained expression on the $m^2/H^2$ series, we get
\begin{align}\label{m_log}
\Bigl\langle \varphi^{2} (t) & \Bigr \rangle = \frac{H^3t}{4\pi^2} \Biggl( \biggl(1-\frac{m^2}{3H^2}\bigl(Ht\bigr) + \frac{2m^4}{27H^4} \bigl(Ht\bigr)^2 - \frac{m^6}{81H^6} \bigl(Ht\bigr)^3 + \, ... \, \biggr) \\ \nonumber
& \qquad \qquad \quad \, - \, \frac{\lambda}{6\pi^2} \bigl(Ht\bigr)^2 \biggl(1-\frac{2m^2}{3H^2}\bigl(Ht\bigr) + \frac{11m^4}{45H^4} \bigl(Ht\bigr)^2 - \frac{26m^6}{405H^6} \bigl(Ht\bigr)^3 + \, ... \, \biggr) \\ \nonumber
& \quad \, + \, \frac{\lambda^2}{20\pi^4} \bigl(Ht\bigr)^4 \biggl(1-\frac{26m^2}{27H^2}\bigl(Ht\bigr) + \frac{40m^4}{81H^4} \bigl(Ht\bigr)^2 - \frac{101m^6}{567H^6} \bigl(Ht\bigr)^3 + \, ... \, \biggr)\Biggr) + O\left(\lambda^3\right).
\end{align} One can trace out the leading infrared logarithm structure in the cosmic time $t$ and establish in the $m^2=0$ case the correspondence to series~\eqref{2pt_massless_series} above and~\cite{2005NuPhB.724..295T,1998PhLB..426...21T,2007PhRvD..76d3512K,2015PhRvD..91j3537O,2020PhRvD.102f5010K,2025JHEP...04..127K}. The same reasoning is valid for the four-point, six-point, and so on correlation functions. Without self-interaction, one has the first line of~\eqref{m_log}, which can also be extracted via the Fokker-Planck equation; see the corresponding appendix in~\cite{2025JHEP...04..127K}. The obtained equal-time correlation function~\eqref{two-point_tt} is the precisely resummed expression of the first line of~\eqref{m_log}.

\section{{Useful integrals}}\label{appendix}
In this short appendix we provide some useful integral's results:
\begin{align}
\int\limits_{-\infty}^{+\infty} dx \, x \, \e^{ \, \alpha xy-\beta x^2} & = \frac{\sqrt{\pi} \, \alpha\, y}{2 \, \beta^{\, 3/2}} \,  \e^{ \, \tfrac{ \, \alpha^2 y^2}{4\beta}} ; \\ 
\int\limits_{-\infty}^{+\infty} dx \, x^2 \, \e^{ \, \alpha xy-\beta x^2} & 
= \frac{\sqrt{\pi} \, }{4 \, \beta^{\, 5/2}} \, \bigl(\alpha^2\, y^2 + 2 \beta \, \bigr) \, \e^{ \, \tfrac{ \, \alpha^2 y^2}{4\beta}} ; \\
\int\limits_{-\infty}^{+\infty} dx \, \bigl(x^3 + \gamma x\bigr) \, \e^{ \, \alpha xy-\beta x^2} &= \frac{\sqrt{\pi} \, \alpha^3}{8 \, \beta^{\, 7/2}} \, \biggl( y^3 +  \frac{2 \beta}{\alpha^2} \bigl( 2  \gamma \beta + 3\bigr) \, y \, \biggr)\, \e^{ \, \tfrac{ \, \alpha^2 y^2}{4\beta}} ; \\
\int\limits_{-\infty}^{+\infty} dx \,\bigl( x^6 + \gamma x^4 \, \bigr)\, \, \e^{ \, - \beta x^2} & = \frac{3 \, \sqrt{\pi}}{8 \, \beta^{\, 7/2}} \, \bigl(2 \gamma \beta +5 \bigr).
\end{align} We have used them to calculate the multi-time correlation functions~\eqref{two-point_t1t2},~\eqref{4-pt_3t1t2_1}, and~\eqref{integrand_four-point_t1t2}. 

\bibliographystyle{JHEP}
\bibliography{biblio.bib}

\providecommand{\href}[2]{#2}\begingroup\raggedright\begin{thebibliography}{10}

\bibitem{1986LNP...246..107S}
A.A.~{Starobinsky}, \emph{{Stochastic de Sitter (inflationary) Stage in the Early Universe}},  in \emph{Field Theory, Quantum Gravity and Strings}, H.J.~{de Vega} and N.~{S{\'a}nchez}, eds., vol.~246, p.~107 (1986), \href{https://doi.org/10.1007/3-540-16452-9_6}{DOI}.

\bibitem{1994PhRvD..50.6357S}
A.A.~{Starobinsky} and J.~{Yokoyama}, \emph{{Equilibrium state of a self-interacting scalar field in the de Sitter background}}, \href{https://doi.org/10.1103/PhysRevD.50.6357}{\emph{Phys. Rev.~D} {\bfseries 50} (1994) 6357} [\href{https://arxiv.org/abs/astro-ph/9407016}{{\ttfamily astro-ph/9407016}}].

\bibitem{Vilenkin:1982wt}
A.~Vilenkin and L.H.~Ford, \emph{{Gravitational Effects upon Cosmological Phase Transitions}}, \href{https://doi.org/10.1103/PhysRevD.26.1231}{\emph{{Phys. Rev. D}} {\bfseries 26} (1982) 1231}.

\bibitem{Linde:1982uu}
A.D.~Linde, \emph{{Scalar Field Fluctuations in Expanding Universe and the New Inflationary Universe Scenario}}, \href{https://doi.org/10.1016/0370-2693(82)90293-3}{\emph{Phys. Lett. B} {\bfseries 116} (1982) 335}.

\bibitem{2005NuPhS.148..108W}
R.P.~{Woodard}, \emph{{A Leading Log Approximation for Inflationary Quantum Field Theory}}, \href{https://doi.org/10.1016/j.nuclphysbps.2005.04.056}{\emph{Nucl. Phys. B Proc. Suppl.} {\bfseries 148} (2005) 108} [\href{https://arxiv.org/abs/astro-ph/0502556}{{\ttfamily astro-ph/0502556}}].

\bibitem{2005NuPhB.724..295T}
N.C.~{Tsamis} and R.P.~{Woodard}, \emph{{Stochastic quantum gravitational inflation}}, \href{https://doi.org/10.1016/j.nuclphysb.2005.06.031}{\emph{Nucl. Phys.~B} {\bfseries 724} (2005) 295} [\href{https://arxiv.org/abs/gr-qc/0505115}{{\ttfamily gr-qc/0505115}}].

\bibitem{2025arXiv250115843W}
R.P.~{Woodard}, \emph{{Recent Developments in Stochastic Inflation}}, \href{https://doi.org/10.48550/arXiv.2501.15843}{\emph{arXiv e-prints} } [\href{https://arxiv.org/abs/2501.15843}{{\ttfamily 2501.15843}}].

\bibitem{2007PhRvD..76d3512K}
E.O.~{Kahya} and V.K.~{Onemli}, \emph{{Quantum stability of a $w < -1$ phase of cosmic acceleration}}, \href{https://doi.org/10.1103/PhysRevD.76.043512}{\emph{Phys. Rev. D} {\bfseries 76} (2007) 043512} [\href{https://arxiv.org/abs/gr-qc/0612026}{{\ttfamily gr-qc/0612026}}].

\bibitem{2013arXiv1305.0229R}
G.~{Rigopoulos}, \emph{{Fluctuation-dissipation and equilibrium for scalar fields in de Sitter}}, \href{https://doi.org/10.48550/arXiv.1305.0229}{\emph{arXiv e-prints} } [\href{https://arxiv.org/abs/1305.0229}{{\ttfamily 1305.0229}}].

\bibitem{2014PhRvD..89f3506G}
B.~{Garbrecht}, G.~{Rigopoulos} and Y.~{Zhu}, \emph{{Infrared correlations in de Sitter space: Field theoretic versus stochastic approach}}, \href{https://doi.org/10.1103/PhysRevD.89.063506}{\emph{Phys. Rev.~D} {\bfseries 89} (2014) 063506} [\href{https://arxiv.org/abs/1310.0367}{{\ttfamily 1310.0367}}].

\bibitem{2015PhRvD..91f3520G}
B.~{Garbrecht}, F.~{Gautier}, G.~{Rigopoulos} and Y.~{Zhu}, \emph{{Feynman diagrams for stochastic inflation and quantum field theory in de Sitter space}}, \href{https://doi.org/10.1103/PhysRevD.91.063520}{\emph{Phys. Rev.~D} {\bfseries 91} (2015) 063520} [\href{https://arxiv.org/abs/1412.4893}{{\ttfamily 1412.4893}}].

\bibitem{2015PhRvD..91j3537O}
V.K.~{Onemli}, \emph{{Vacuum fluctuations of a scalar field during inflation: Quantum versus stochastic analysis}}, \href{https://doi.org/10.1103/PhysRevD.91.103537}{\emph{Phys. Rev. D} {\bfseries 91} (2015) 103537} [\href{https://arxiv.org/abs/1501.05852}{{\ttfamily 1501.05852}}].

\bibitem{2017JHEP...11..065C}
H.~{Collins}, R.~{Holman} and T.~{Vardanyan}, \emph{{The quantum Fokker-Planck equation of stochastic inflation}}, \href{https://doi.org/10.1007/JHEP11(2017)065}{\emph{JHEP} {\bfseries 2017} (2017) 65} [\href{https://arxiv.org/abs/1706.07805}{{\ttfamily 1706.07805}}].

\bibitem{2018PhRvD..97l3531K}
G.~{Karakaya} and V.K.~{Onemli}, \emph{{Quantum effects of mass on scalar field correlations, power spectrum, and fluctuations during inflation}}, \href{https://doi.org/10.1103/PhysRevD.97.123531}{\emph{Phys. Rev. D} {\bfseries 97} (2018) 123531} [\href{https://arxiv.org/abs/1710.06768}{{\ttfamily 1710.06768}}].

\bibitem{2020JHEP...07..119B}
M.~{Baumgart} and R.~{Sundrum}, \emph{{De Sitter diagrammar and the resummation of time}}, \href{https://doi.org/10.1007/JHEP07(2020)119}{\emph{JHEP} {\bfseries 2020} (2020) 119} [\href{https://arxiv.org/abs/1912.09502}{{\ttfamily 1912.09502}}].

\bibitem{2020PhRvD.102f5010K}
A.Y.~{Kamenshchik} and T.~{Vardanyan}, \emph{{Renormalization group inspired autonomous equations for secular effects in de Sitter space}}, \href{https://doi.org/10.1103/PhysRevD.102.065010}{\emph{Phys. Rev. D} {\bfseries 102} (2020) 065010} [\href{https://arxiv.org/abs/2005.02504}{{\ttfamily 2005.02504}}].

\bibitem{2022EPJC...82..345K}
A.Y.~{Kamenshchik}, A.A.~{Starobinsky} and T.~{Vardanyan}, \emph{{Massive scalar field in de Sitter spacetime: a two-loop calculation and a comparison with the stochastic approach}}, \href{https://doi.org/10.1140/epjc/s10052-022-10295-z}{\emph{Eur. Phys. J. C} {\bfseries 82} (2022) 345} [\href{https://arxiv.org/abs/2109.05625}{{\ttfamily 2109.05625}}].

\bibitem{2022JHEP...02..121A}
J.O.~{Andersen}, M.~{Eriksson} and A.~{Tranberg}, \emph{{Stochastic inflation from quantum field theory and the parametric dependence of the effective noise amplitude}}, \href{https://doi.org/10.1007/JHEP02(2022)121}{\emph{JHEP} {\bfseries 2022} (2022) 121} [\href{https://arxiv.org/abs/2111.14503}{{\ttfamily 2111.14503}}].

\bibitem{2023JCAP...03..058B}
S.~{Bhattacharya} and N.~{Joshi}, \emph{{Non-perturbative analysis for a massless minimal quantum scalar with V({\ensuremath{\phi}}) = {\ensuremath{\lambda}}{\ensuremath{\phi}} $^{4}$/4! + {\ensuremath{\beta}}{\ensuremath{\phi}} $^{3}$/3! in the inflationary de Sitter spacetime}}, \href{https://doi.org/10.1088/1475-7516/2023/03/058}{\emph{JCAP} {\bfseries 2023} (2023) 058} [\href{https://arxiv.org/abs/2211.12027}{{\ttfamily 2211.12027}}].

\bibitem{2023JHEP...08..060H}
M.~{Honda}, R.~{Jinno}, L.~{Pinol} and K.~{Tokeshi}, \emph{{Borel resummation of secular divergences in stochastic inflation}}, \href{https://doi.org/10.1007/JHEP08(2023)060}{\emph{JHEP} {\bfseries 2023} (2023) 60} [\href{https://arxiv.org/abs/2304.02592}{{\ttfamily 2304.02592}}].

\bibitem{2024JHEP...04..004C}
S.~{C{\'e}spedes}, A.-C.~{Davis} and D.-G.~{Wang}, \emph{{On the IR divergences in de Sitter space: loops, resummation and the semi-classical wavefunction}}, \href{https://doi.org/10.1007/JHEP04(2024)004}{\emph{JHEP} {\bfseries 2024} (2024) 4} [\href{https://arxiv.org/abs/2311.17990}{{\ttfamily 2311.17990}}].

\bibitem{1998PhLB..426...21T}
N.C.~{Tsamis} and R.P.~{Woodard}, \emph{{Matter contributions to the expansion rate of the universe}}, \href{https://doi.org/10.1016/S0370-2693(98)00159-2}{\emph{Phys. Lett.~B} {\bfseries 426} (1998) 21} [\href{https://arxiv.org/abs/hep-ph/9710466}{{\ttfamily hep-ph/9710466}}].

\bibitem{Yang:1950vi}
C.-N.~Yang and D.~Feldman, \emph{{The S Matrix in the Heisenberg Representation}}, \href{https://doi.org/10.1103/PhysRev.79.972}{\emph{Phys. Rev.} {\bfseries 79} (1950) 972}.

\bibitem{2025JHEP...04..127K}
A.~Kamenshchik and P.~Petriakova, \emph{{IR finite correlation functions in de Sitter space, a smooth massless limit, and an autonomous equation}}, \href{https://doi.org/10.1007/JHEP04(2025)127}{\emph{JHEP} {\bfseries 04} (2025) 127} [\href{https://arxiv.org/abs/2410.16226}{{\ttfamily 2410.16226}}].

\bibitem{Uhlenbeck:1930zz}
G.E.~Uhlenbeck and L.S.~Ornstein, \emph{{On the Theory of the Brownian Motion}}, \href{https://doi.org/10.1103/PhysRev.36.823}{\emph{Phys. Rev.} {\bfseries 36} (1930) 823}.

\bibitem{1989fpem.book.....R}
H.~{Risken}, \emph{{The Fokker-Planck equation. Methods of solution and applications}} (1989).

\bibitem{2000PhRvD..62l4019A}
P.R.~{Anderson}, W.~{Eaker}, S.~{Habib}, C.~{Molina-Par{\'\i}s} and E.~{Mottola}, \emph{{Attractor states and infrared scaling in de Sitter space}}, \href{https://doi.org/10.1103/PhysRevD.62.124019}{\emph{Phys. Rev.~D} {\bfseries 62} (2000) 124019} [\href{https://arxiv.org/abs/gr-qc/0005102}{{\ttfamily gr-qc/0005102}}].

\bibitem{2013PhLB..727..541G}
F.~{Gautier} and J.~{Serreau}, \emph{{Infrared dynamics in de Sitter space from Schwinger-Dyson equations}}, \href{https://doi.org/10.1016/j.physletb.2013.10.072}{\emph{Phys. Lett. B} {\bfseries 727} (2013) 541} [\href{https://arxiv.org/abs/1305.5705}{{\ttfamily 1305.5705}}].

\bibitem{2017JCAP...10..018H}
R.J.~{Hardwick}, V.~{Vennin}, C.T.~{Byrnes}, J.~{Torrado} and D.~{Wands}, \emph{{The stochastic spectator}}, \href{https://doi.org/10.1088/1475-7516/2017/10/018}{\emph{JCAP} {\bfseries 2017} (2017) 018} [\href{https://arxiv.org/abs/1701.06473}{{\ttfamily 1701.06473}}].

\bibitem{1983PhRvD..27.2848V}
A.~{Vilenkin}, \emph{{Birth of inflationary universes}}, \href{https://doi.org/10.1103/PhysRevD.27.2848}{\emph{Phys. Rev.~D} {\bfseries 27} (1983) 2848}.

\bibitem{SHAPIRO1978563}
V.~Shapiro and V.~Loginov, \emph{"formulae of differentiation" and their use for solving stochastic equations}, \href{https://doi.org/https://doi.org/10.1016/0378-4371(78)90198-X}{\emph{Phys. A: Stat. Mech. Appl.} {\bfseries 91} (1978) 563}.

\bibitem{1992sppc.book.....V}
N.G.~{van Kampen}, \emph{{Stochastic Processes in Physics and Chemistry}} (1992).

\bibitem{etde_634926}
V.M.~Loginov, \emph{Simple mathematical tool for statistical description of dynamical systems under random actions}, \href{https://doi.org/https://www.actaphys.uj.edu.pl/R/27/3/693/pdf}{\emph{Acta Phys. Pol. B} {\bfseries 27} (1996) 693}.

\end{thebibliography}\endgroup
\end{document}